\documentclass[modern]{aastex62}
\usepackage{natbib}

\newcommand{\sourcefull}{4U 1820$-$30}
\newcommand{\source}{4U 1820}

\shorttitle{X-ray burst winds in \sourcefull}
\shortauthors{Strohmayer et al.}

\begin{document}

\title{NICER Discovers Spectral Lines During Photospheric Radius
  Expansion Bursts from 4U 1820$-$30: Evidence for Burst-driven Winds}

\author[0000-0001-7681-5845]{T. E. Strohmayer} 
\affil{Astrophysics Science Division and Joint Space-Science Institute,
  NASA's Goddard Space Flight Center, Greenbelt, MD 20771, USA}

\author[0000-0002-3422-0074]{D. Altamirano}
\affil{Physics \& Astronomy, University of Southampton, Southampton,
 Hampshire SO17 1BJ, UK}

\author{Z. Arzoumanian} 
\affil{X-ray Astrophysics Laboratory,
 Astrophysics Science Division, NASA's Goddard Space Flight Center,
  Greenbelt, MD 20771, USA} 


\author{P. M. Bult} 
\affil{Astrophysics Science Division, NASA's Goddard
  Space Flight Center, Greenbelt, MD 20771, USA}

\author[0000-0001-8804-8946]{D. Chakrabarty} 
\affil{MIT Kavli Institute for Astrophysics
 and Space Research, Massachusetts Institute of Technology,
  Cambridge, MA 02139, USA}

\author[0000-0002-4397-8370]{J. Chenevez} 
\affil{DTU Space—National Space Institute,
 Technical University of Denmark, Elektrovej 327-328, DK-2800 Lyngby,
  Denmark}

\author{A. C. Fabian}
\affil{Institute of Astronomy, Madingley Road, Cambridge CB3 0HA, UK}


\author{K. C. Gendreau}
\affil{X-ray Astrophysics Laboratory,
 Astrophysics Science Division, NASA's Goddard Space Flight Center,
 Greenbelt, MD 20771, USA} 

\author[0000-0002-6449-106X]{S. Guillot} 
\affil{CNRS, IRAP, 9 avenue du Colonel Roche, BP
  44346, F-31028 Toulouse Cedex 4, France} 
\affil{Universit\'e de Toulouse, CNES, UPS-OMP, F-31028 Toulouse, France}





\author{J. J. M. in 't Zand} 
\affil{SRON Netherlands Institute for
  Space Research, Sorbonnelaan 2, 3584 CA Utrecht, The Netherlands}

\author[0000-0002-6789-2723]{G. K. Jaisawal}
\affil{DTU Space—National Space Institute,
 Technical University of Denmark, Elektrovej 327-328, DK-2800 Lyngby,
  Denmark}

\author{L. Keek} 
\affil{Department of Astronomy, University of Maryland College Park, 
MD 20742, USA}

\author{P. Kosec}
\affil{Institute of Astronomy, Madingley Road, Cambridge CB3 0HA, UK}


\author{R. M. Ludlam} 
\affil{Department of Astronomy, University of
  Michigan, 1085 South University Avenue, Ann Arbor, MI 48109-1107,
  USA}

\author{S. Mahmoodifar} 
\affil{Department of Astronomy, University of Maryland College Park, 
MD 20742, USA}



\author[0000-0002-0380-0041]{Christian Malacaria} \affiliation{NASA
  Marshall Space Flight Center, NSSTC, 320 Sparkman Drive, Huntsville,
  AL 35805, USA}\thanks{NASA Postdoctoral Fellow}
\affiliation{Universities Space Research Association, NSSTC, 320
  Sparkman Drive, Huntsville, AL 35805, USA}

\author{J. M. Miller}
\affil{Department of Astronomy, University of Michigan, 1085 South University
Ave, Ann Arbor, MI 48109-1107, USA}








\begin{abstract}

We report the discovery with the {\it Neutron Star Interior
  Composition Explorer} ({\it NICER}) of narrow emission and
absorption lines during photospheric radius expansion (PRE) X-ray
bursts from the ultracompact binary \sourcefull{}.  {\it NICER}
observed 4U 1820$-$30 in 2017 August during a low flux, hard spectral
state, accumulating about 60 ks of exposure. Five thermonuclear X-ray
bursts were detected of which four showed clear signs of PRE.  We
extracted spectra during the PRE phases and fit each to a model that
includes a comptonized component to describe the accretion-driven
emission, and a black body for the burst thermal radiation. The
temperature and spherical emitting radius of the fitted black body are
used to assess the strength of PRE in each burst.  The two strongest
PRE bursts (burst pair 1) had black body temperatures of
$\approx{0.6}$ keV and emitting radii of $\approx 100$ km (at a
distance of 8.4 kpc). The other two bursts (burst pair 2) had higher
temperatures ($\approx 0.67$ keV) and smaller radii ($\approx 75$
km). All of the PRE bursts show evidence of narrow line emission near
1 keV. By co-adding the PRE phase spectra of burst pairs 1 and,
separately, 2 we find, in both co-added spectra, significant, narrow,
spectral features near 1.0 (emission), 1.7 and 3.0 keV (both in
absorption). Remarkably, all the fitted line centroids in the co-added
spectrum of burst pair 1 appear systematically blue-shifted by a
factor of $1.046 \pm 0.006$ compared to the centroids of pair 2,
strongly indicative of a gravitational shift, a wind-induced
blue-shift, or more likely some combination of both effects. The
observed shifts are consistent with this scenario in that the stronger
PRE bursts in pair 1 reach larger photospheric radii, and thus have
weaker gravitational redshifts, and they generate faster outflows,
yielding higher blue-shifts.  We discuss possible elemental
identifications for the observed features in the context of recent
burst-driven wind models.


\end{abstract} 
\keywords{stars: neutron --- X-rays: binaries --- X-rays: bursts ---
  X-rays: individual (4U 1820$-$30) --- radiation: dynamics}

\section{Introduction}
\label{sec:introduction} 

Thermonuclear X-ray bursts provide unique opportunities to study the
extreme physics of neutron stars.  These bright X-ray events result
from unstable burning in an accreted layer of light elements on the
neutron star's surface. The strongly heated surface layers briefly
shine brightly in the X-ray band, and as the radiation comes directly
from the stellar surface, it provides important clues regarding
neutron star structure and the nuclear burning on them (see
\citealt{2006csxs.book..113S, 2017arXiv171206227G}, for reviews).  The
continuum emission during bursts is broadly consistent with black body
emission from the heated surface layers. A major focus of recent
efforts has been to employ the observed spectra to constrain the
stellar mass and radius \citep{1993SSRv...62..223L}.  The basic idea
is to measure an X-ray flux, $F_x$, and color temperature, $T_{\rm
  col}$, and then use an appropriate atmosphere model to convert the
color temperature to an effective temperature, $T_{\rm eff}$. One can
then, in principle, infer an assumed spherical emission area from the
Stefan Boltzmann law, $L = 4\pi d^2 F_x = 4\pi R^2 \sigma T_{\rm
  eff}^4$, where $d$ and $R$ are the source distance and photospheric
radius, respectively. For bursts that reach the Eddington luminosity,
an additional constraint is available, since this quantity depends
only on the stellar mass, radius, and the composition of the
atmosphere.  A full treatment must also account for the significant
gravitational redshift from the neutron star surface, as well as any
special relativistic effects associated with the potentially rapid
spin of its surface.  Substantial efforts have been made by a number
of research groups over the past decade to carry out this program
(see, for example, \citealt{2006Natur.441.1115O, 2010ApJ...722...33S,
  2017A&A...608A..31N, 2016ApJ...829...48G}, and references therein).
While very promising, significant uncertainties remain in interpreting
such observations.  These can be illustrated with a few rhetorical
questions: how accurate is the atmosphere model, in particular its
composition?, are other emission components, for example, reprocessed
emission from an accretion disk, present in the observed spectrum?,
how accurately are the distance, flux, and color temperature known?.

Another way to obtain constraints on the stellar mass and radius is to
infer a surface gravitational redshift, $z = (E_0 - E)/E$, by
observing a spectral feature; a line or edge at energy $E$ that
originates from the stellar surface, of a known, rest-frame transition
with energy $E_0$. A constraint on the mass to radius ratio, the
``compactness,'' $M/R$, then follows from the relation $1+z = (1 -
2GM/c^2 R)^{-1/2}$. While perhaps conceptually simpler than correctly
interpreting observed continuum spectra, difficulties remain,
including; is the line identification secure?, is the emission site at
the stellar surface, or some other altitude or location?  Perhaps most
fundamental of all is that it has been very difficult to reliably
identify {\it any} surface spectral features in X-ray burst spectra.

There were some early claims of absorption line detections from the
bursters 4U 1636$-$536, 4U 1608$-$522, and EXO 1747$-$214 based on
data from {\it Tenma} and {\it EXOSAT} \citep{1984PASJ...36..819W,
  1988PASJ...40..209N, 1989MNRAS.237..729M}, but subsequent
observations with {\it ASCA}, {\it RXTE} and {\it BeppoSAX} did not
tend to confirm these initial reports.  There has also been a recent
report based on {\it NuSTAR} observations of GRS 1741.9$-$2853 of an
absorption line at $5.46 \pm 0.1$ keV during an X-ray burst from this
object \citep{2015ApJ...799..123B}. However, the reported significance
of the line was rather modest ($1.7 \sigma$).  Observations of two
superbursts, from 4U 1820$-$30 and 4U 1636$-$536, with the {\it RXTE}
Proportional Counter Array (PCA) clearly show emission features in the
Fe K band that have been succesfully modelled as burst thermal
emission reprocessed (reflected) by the accretion disk
\citep{2002ApJ...566.1045S, 2014ApJ...789..121K,
  2014ApJ...797L..23K}. While these features provide an exciting view
of the interaction of the burst flux with the accretion disk
\citep{2004ApJ...602L.105B, 2018ApJ...867L..28F}, it seems highly
likely that they do not originate at the stellar surface.

\cite{2002Natur.420...51C} reported on a set of narrow absorption
lines in the co-added, high resolution spectra of 28 bursts from EXO
0748$-$676 observed with the {\it XMM-Newton} Reflection Grating
Spectrometer (RGS).  They argued that these features could be
associated with transitions of Fe XXV, Fe XXVI, and O VIII at a
consistent surface redshift of $1 + z = 1.35$. Unfortunately,
subsequent observations did not confirm these lines, and the likely
stellar spin frequency was determined to be 552 Hz
\citep{2010ApJ...711L.148G}, substantially higher than the 45 Hz value
that was initially reported for the source
\citep{2004ApJ...614L.121V}.  The higher spin rate implies rotational
broadening that would be incompatible with the narrow absorption lines
initially claimed \citep{2010ApJ...723.1053L}.  Additionally, detailed
model atmosphere calculations by \cite{2008A&A...490.1127R} did not
confirm the line identifications proposed by
\cite{2002Natur.420...51C}.  Whether the lines are real or not remains
controversial.

More recently, there have been reports of absorption edges being
detected during X-ray bursts.  \cite{2010A&A...520A..81I} reported
evidence from {\it RXTE/PCA} data for edges during so-called
``superexpansion'' bursts from 4U 0614+091, 4U 1722$-$30, and 4U
1820$-$30. These are bursts with powerful photospheric radius
expansion (PRE) phases, that very likely drive outflows
\citep{2012A&A...547A..47I, 2018ApJ...863...53Y}.  The edge energies
were in the 5 - 12 keV range, and were detected after the strong,
initial expansion phase.  \cite{2013ApJ...767L..37D} reported the
detection with the {\it Swift/XRT} of a 1 keV emission line as well as
Fe-K band absorption features (lines and edges) in an energetic,
intermediate-duration thermonuclear burst from the ultracompact
accreting millisecond X-ray pulsar (AMXP) IGR J17062$-$6143
\citep{2018ApJ...858L..13S}.  They suggested the 1 keV emission line
could be associated with Fe-L shell transitions from the irradiation
of cold gas by the burst, or alternatively, the $1s$ - $2p$
Ly-$\alpha$ transition of Ne X could also account for the observed
line energy.

\cite{2017MNRAS.464L...6K} report evidence for absorption edges at
decreasing energies between 8 and 7.6 keV in a powerful He flash from
the AMXP HETE J1900.1$-$2455. Like \cite{2010A&A...520A..81I}, they
argue that these features are associated with metals--the ashes of the
thermonuclear burning--that have been dredged up by the vigorous
convection initiated by such bursts \citep{2006ApJ...639.1018W}.
\cite{2017MNRAS.464L...6K} identify the features in HETE
J1900.1$-$2455 with the 9.28 keV photoionization edge of hydrogen-like
Fe ions, but redshifted from the stellar surface to the observed
energies.  The variation in observed edge energy is attributed to
variation in the vertical location of the photosphere.  Based on this
the authors estimate $1+z \approx 1.2$, which is roughly consistent
with the redshift expected from a neutron star with $M = 1.4
M_{\odot}$ and $R = 12$ km.  \cite{2018ApJ...866...53L} also claim an
absorption edge is present during a PRE burst from GRS 1747$-$312
observed with the {\it RXTE/PCA}. They report evolution of the
absorption edge energy from 9.45 to $\approx 6$ keV through the PRE
phase, and then to $\approx 8$ keV in the cooling tail.  They
attribute the shift from 9.45 to 8 keV to gravitational redshift of
the hydrogen-like Ni photoionization edge between the neutron star
surface and the expanded photosphere during PRE. We note that this is
the same burst for which \cite{2003A&A...409..659I} reported evidence
of a broad emission feature at 4.8 keV during the PRE phase, so
interpretation of the apparent features in this burst as edges may not
be unique.

Many of the reported detections of spectral features in X-ray bursts
to date have been made with data from instruments with relatively poor
resolution, such as {\it RXTE's} PCA, which has $\approx 20 \%$
resolution at 6 keV.  Higher resolution detectors have typically
suffered from some technical problems when observing bright X-ray
bursts. The high count rates mean that CCD detectors suffer pile-up or
the relatively higher backgrounds present when they must be clocked at
high rates.  Whereas dispersive spectroscopy with either {\it Chandra}
or {\it XMM-Newton} generally mitigates any pile-up, these instruments
bring relatively modest effective area to bear on the brief intervals
for which bursts are bright.  Fortunately, {\it NICER's} X-ray Timing
Instrument (XTI) brings CCD-like spectral resolution ($\approx 80$ eV
at 1 keV), and substantial, pile-up-free effective area to
observations of X-ray bursts.  Moreover, {\it NICER's} bandpass
extends down to 0.2 keV, which is ideal for observing the soft-X-ray
expansion phases of PRE bursts. With the recently reported detections
of edge features in such bursts, we aim to leverage {\it NICER}'s
unique capabilities to search for similar features.

In this paper we report results of {\it NICER} observations of
thermonuclear X-ray bursts from 4U 1820$-$30 (hereafter, 4U 1820)
which reveal both emission and absorption features during their PRE
phases. \source{} is a well-known, ultracompact accreting neutron star
binary with an 11.4 min orbital period. The system was one of the
first discovered to produce thermonuclear X-ray bursts
\citep{1976ApJ...205L.127G}. Its compact nature requires a low-mass
degenerate dwarf donor \citep{1987ApJ...322..842R}, and helium is
likely the most abundant burst fuel \citep{1987ApJ...314..266H,
  1995ApJ...438..852B, 2003ApJ...595.1077C}.

The paper is organized as follows.  We first briefly describe the
observations made and the general properties of the observed bursts.
We then discuss our spectral analysis and the evidence for an emission
line at 1 keV, and absorption lines at 1.7 and 3.0 keV in the PRE
phases of these bursts. We also show that there is a systematic energy
shift present in the co-added spectra of the pair of strongest PRE
bursts, compared to those in the burst pair with weaker PRE.  We then
discuss several scenarios for the origin of the line features, and
provide possible elemental identifications. We also explore the idea
that some of the observed energy shift is related to a gravitational
redshift, a wind-induced blue-shift, or perhaps both. We conclude with
a brief summary and discussion of the implications for models of burst
winds and future efforts to probe their composition, energetics and
formation.

\section{NICER Observations}

\label{sec:observations} 
{\it NICER} is an X-ray observatory operating on the International
Space Station ({\it ISS}). It provides low background, high throughput
($\approx 1900$ cm$^2$ at 1.5 keV), high time resolution observations
across the $0.2 - 12$ keV X-ray band \citep{2012SPIE.8443E..13G}.
{\it NICER} observed \source{} in 2017 August, accumulating about 60
ks of exposure as part of a science team program with a major goal to
detect X-ray bursts and search for spectral features in them. These
observations were triggered by the entry of the source into a low flux
and hard spectral state in which it is known to produce bursts
\citep{2018ApJ...856L..37K}. These data were processed using the
NICERDAS (V004) software, and standard filtering and cleaning were
applied.  This involves excluding all data where the pointing offset
is $> 54$ arcsec, the dark Earth limb angle is $< 30^{\circ}$, the
bright Earth limb angle is $< 40^{\circ}$, and the ISS was inside the
South Atlantic Anomaly (SAA). We employed version v1.02 of the {\it
  NICER} response functions in our analysis.

\subsection{Burst Light Curves}

\begin{figure*}
\begin{center}
\includegraphics[scale=0.75]{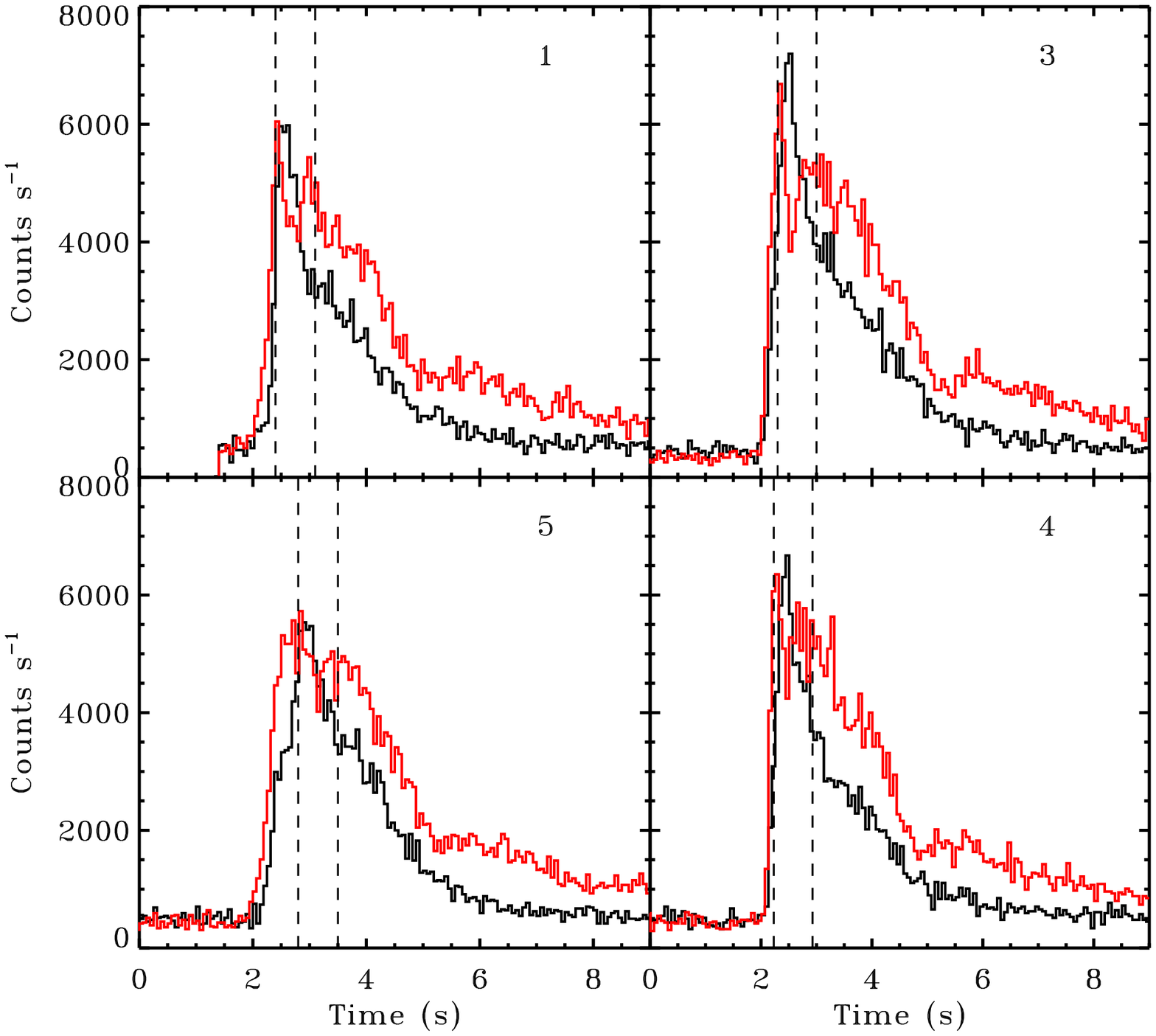}%
\end{center}
\caption{\label{fig:lc1} Light curves of four PRE X-ray bursts from
  \source{} observed with {\it NICER}. Count rates computed in 1/16 s
  intervals in the 0.2 - 1.0 keV (black) and 2.0 - 10.0 keV (red)
  bands are shown. The vertical dotted lines mark the intervals used
  to extract PRE-phase spectra. Clockwise from upper left, the four
  panels correspond to bursts 1, 3, 4, and 5 (see Table
  \ref{table:bursts}). }
\end{figure*}

\begin{figure*}
\begin{center}
\includegraphics[scale=0.75]{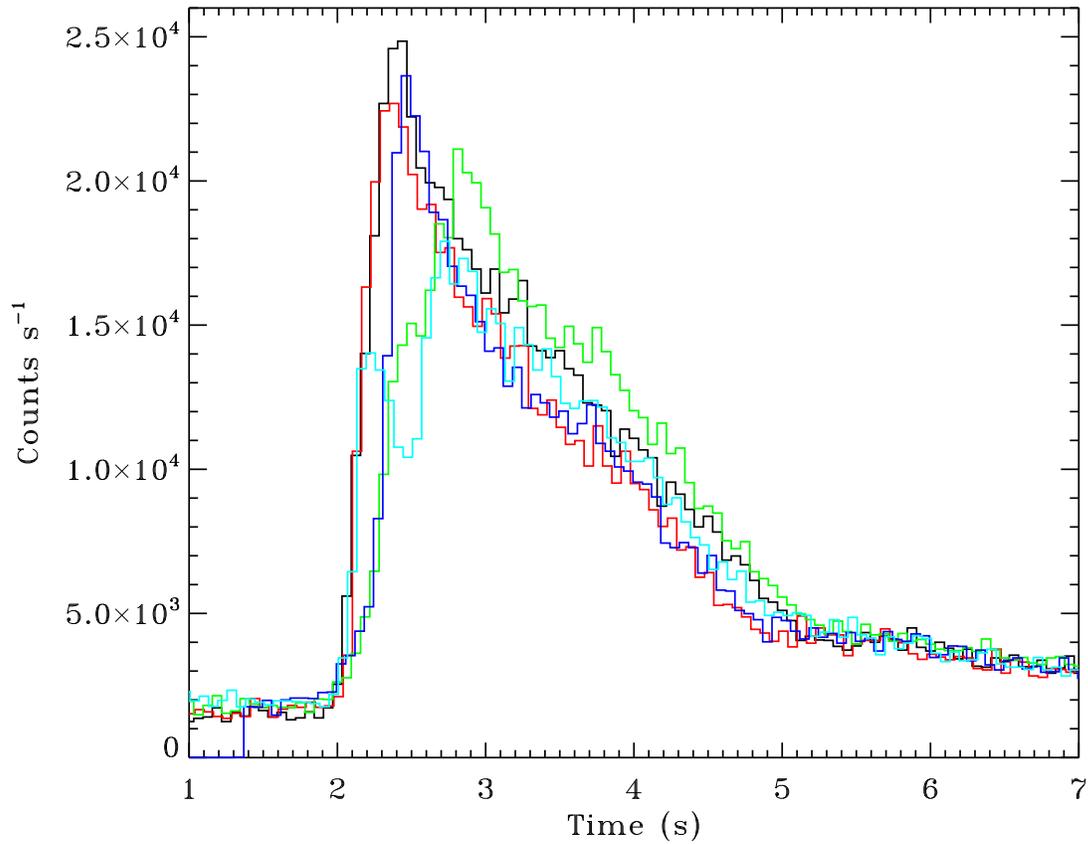}%
\end{center}
\caption{\label{fig:lc2} Light curves of all five X-ray bursts from
  \source{} observed with {\it NICER}. Count rates computed in 1/16 s
  intervals in the full {\it NICER} band (0.2 -10.0 keV) are plotted.
  For comparison purposes, the start times have been approximately
  aligned.  The different colors correspond to the following bursts; 1
  (blue), 2 (cyan), 3 (black), 4 (red), and 5 (green).  Burst 2 (cyan)
  is ``anomalous" in that it shows a significant drop in count rate in
  both the soft and hard bands, and does not show evidence of PRE. }

\end{figure*}

Five bursts were observed during this campaign (see Table
\ref{table:bursts}). We note that detailed, time resolved spectroscopy
of burst 1 was reported in \cite{2018ApJ...856L..37K}, and it is not
our goal here to reproduce such an analysis for all the observed
bursts.  We began by extracting light curves for each burst in several
energy bands; 0.2 - 1 keV (soft), 2 - 10 keV (hard), and the full band
0.2 - 10 keV.  Figure \ref{fig:lc1} shows the resulting soft and hard
band light curves for bursts 1 (top left), 3 (top right), 4 (bottom
right), and 5 (bottom left), all of which show evidence for PRE, in
that the soft band rate (black curves) peaks while there is a
correlated drop in the hard band rate (red curves).  Burst 2 is
``anomalous'' in this regard. The hard band rate is actually larger
than that in the soft band, and there are correlated drops in both
rates during the burst rise. This behavior is suggestive of an
additional absorption component present during this burst.  Because it
lacks a clear signature of PRE we did not consider it in the spectral
analysis reported here. We will explore this interesting burst
elsewhere.  In Figure \ref{fig:lc2} we show the light curves of all
five bursts in the full band (0.2 - 10 keV). The caption gives the
color code to identify the bursts. We approximately aligned the rise
times for display purposes and so that the profiles can be compared.
The cyan curve is burst 2, and the prominent dip after the initial
rise is apparent.

\subsection{Single Burst Spectra}

Guided by the results of \cite{2018ApJ...856L..37K} we began by
extracting spectra during the PRE phases for each burst.  We
identified the interval where the 2 - 10 keV band light curve peaks
and extracted spectra for 0.7 s from this point for each burst. This
exposure value approximately encompasses the lowest temperature
portion of the PRE phase (see \cite{2018ApJ...856L..37K}, Fig. 3), and
these intervals are marked in Figure \ref{fig:lc1} by the vertical
dashed lines.  We regrouped the spectra by combining every three
successive, native energy (PI) channels, and we also applied a
normalization factor derived from spectral fits to data from the Crab
nebula.  See \cite{2018ApJ...858L...5L} for a discussion of this
procedure, which provides a way to reduce the influence of the
strongest remaining unmodeled residuals in the {\it NICER} response
function. We then fit each spectrum in the band 0.3 - 7 keV with a two
component model comprising a Comptonized component for the persistent
emission and a black body for the thermal burst flux. We note that the
{\it NICER} background is always negligible in this energy band for
these bright source spectra.  In XSPEC this model has the form, {\it
  TBabs*(nthcomp + bbodyrad)}.  See \cite{1999MNRAS.309..561Z} for a
description of the {\it nthcomp} model.

The parameters of the {\it nthcomp} (persistent) component are
qualitatively similar for all bursts. For example, for burst 3 we
find, $\Gamma = 2.2 \pm 0.1$, $kT_{bb} = 0.14 \pm 0.03$ keV, and a
normalization of $7.9 \pm 1.1$ (errors are $1\sigma$). The
normalization is defined as the photon flux (with units, cm$^{-2}$
s$^{-1}$ keV$^{-1}$) at 1 keV.  These spectra are not sensitive to the
electron temperature, $kT_{e}$, in this model, so for all fits we held
it fixed at $8$ keV.  We left all other parameters describing the {\it
  nthcomp} component free to vary, but we note that the fitted
parameter values describing the spectral shape are all qualitatively
similar to those derived from fitting the pre-burst (persistent)
emission with the same model.  The derived black body temperature and
spherical emitting radius characterizing the PRE phase do show
significant variations burst to burst, and these are listed in Table
\ref{table:bursts} for each burst.

\begin{figure*}
\begin{center}
\includegraphics[scale=0.8]{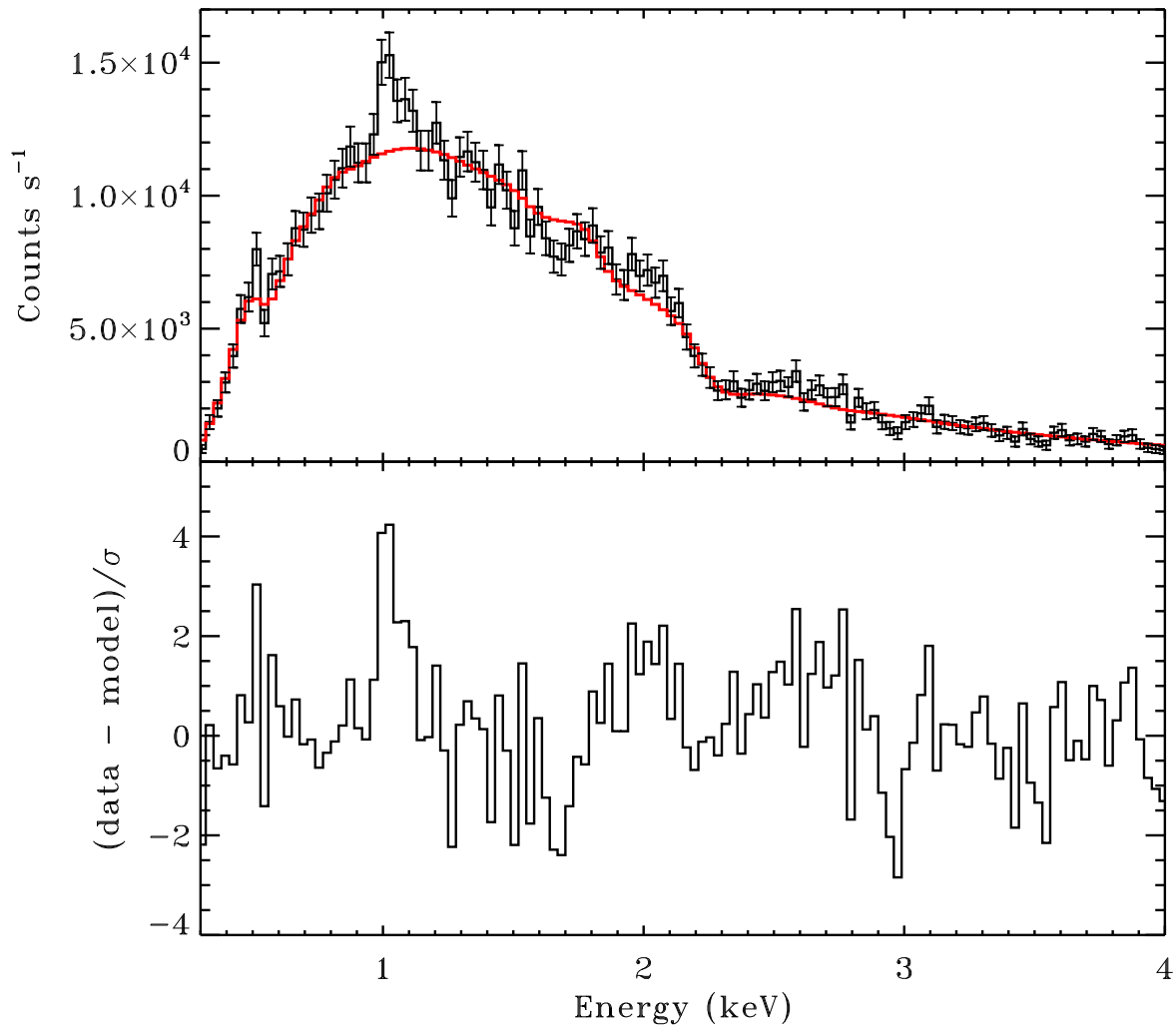}%
\end{center}
\caption{\label{fig:b3spec} X-ray count rate spectrum extracted during
  the PRE phase for burst 3. The top panel shows the observed count
  rate as a function of energy. The red curve is the best-fitting
  continuum model.  The fit residuals, in units of (data -
  model)/$\sigma$, are shown in the bottom panel.  Excess emission is
  evident near 1 keV.  See \S 2.2 for additional details.}

\end{figure*}

The above continuum model provides a reasonable, qualitative fit to
the PRE phase spectra, but there are substantial residuals evident in
the spectra.  The most prominent of these is an excess near 1 keV in
all of the bursts, but it is most obvious in bursts 1 and 3.  As an
example, we show in Figure \ref{fig:b3spec} the PRE phase spectrum
extracted for burst 3, where one can see excess emission above the
continuum near 1 keV.  Adding a gaussian component to model this
feature results in an improvement in $\chi^2$ of 34.8. The fitted line
centroid is $1.04 \pm 0.01 $ keV, and an estimate of the significance,
defined as the ratio of the line normalization to its $1 \sigma$
uncertainty, is $6.1$. The equivalent width is 47 eV.  The line is
unresolved in the sense that we only find an upper limit to its
intrinsic width of $\approx 70$ eV ($90\%$ confidence).  The observed
width, however, is consistent with {\it NICER's} measured spectral
resolution at 1 keV.

\begin{table*}
\renewcommand{\arraystretch}{1.1}
\caption{X-ray bursts from \source{} observed by {\it NICER}}
\label{table:bursts}
\scalebox{0.95}{
\begin{tabular}{cccccc}
\tableline\tableline
Number & ObsID & Start Time & Date & $kT_{PRE}$ & $R_{PRE}$ \\
       &       & MJD (TT) & YYYY-MM-DD & (keV) & (km)  \\
\tableline
1  & 1050300108 & 57994.37115 & 2017-08-29 & $0.573 \pm 0.025$ & $103 \pm 9$ \\
2  & 1050300108 & 57994.46169 & 2017-08-29 & NA & NA  \\
3  & 1050300109 & 57995.22281 & 2017-08-30 & $0.619 \pm 0.026$ & $102 \pm 7$ \\
4  & 1050300109 & 57995.33890 & 2017-08-30 & $0.675 \pm 0.030$ & $78  \pm 6$ \\
5  & 1050300109 & 57995.60332 & 2017-08-30 & $0.664 \pm 0.035$ & $72 \pm 7$ \\
\tableline
\end{tabular}}
\tablecomments{Assigned number, observation ID, burst start time, date
  of observation, black body temperature (keV), and derived
  photospheric radius (km, at 8.4 kpc).}
\end{table*}

\subsection{Co-added Spectra}

As noted above, a similar 1 keV excess is apparent in the PRE spectra
from each of the bursts, but with somewhat varying strength. Motivated
by this we co-added bursts in pairs to improve the signal to noise
ratio.  In order to combine bursts that are most similar to one
another, we were guided by their spectral properties inferred during
the PRE phase (see Table \ref{table:bursts}). The bursts with the
lowest PRE temperatures and the largest photospheric radii are bursts
1 and 3.  Bursts 4 and 5 have higher temperatures and smaller radii
that are consistent within the uncertainties.  The burst pairs that
are most similar to each other are therefore bursts 1 and 3 (pair 1),
and bursts 4 and 5 (pair 2). We thus created two new spectra by
co-adding the single-burst spectra within each pair.

\begin{figure*}
\begin{center}
\includegraphics[scale=0.75]{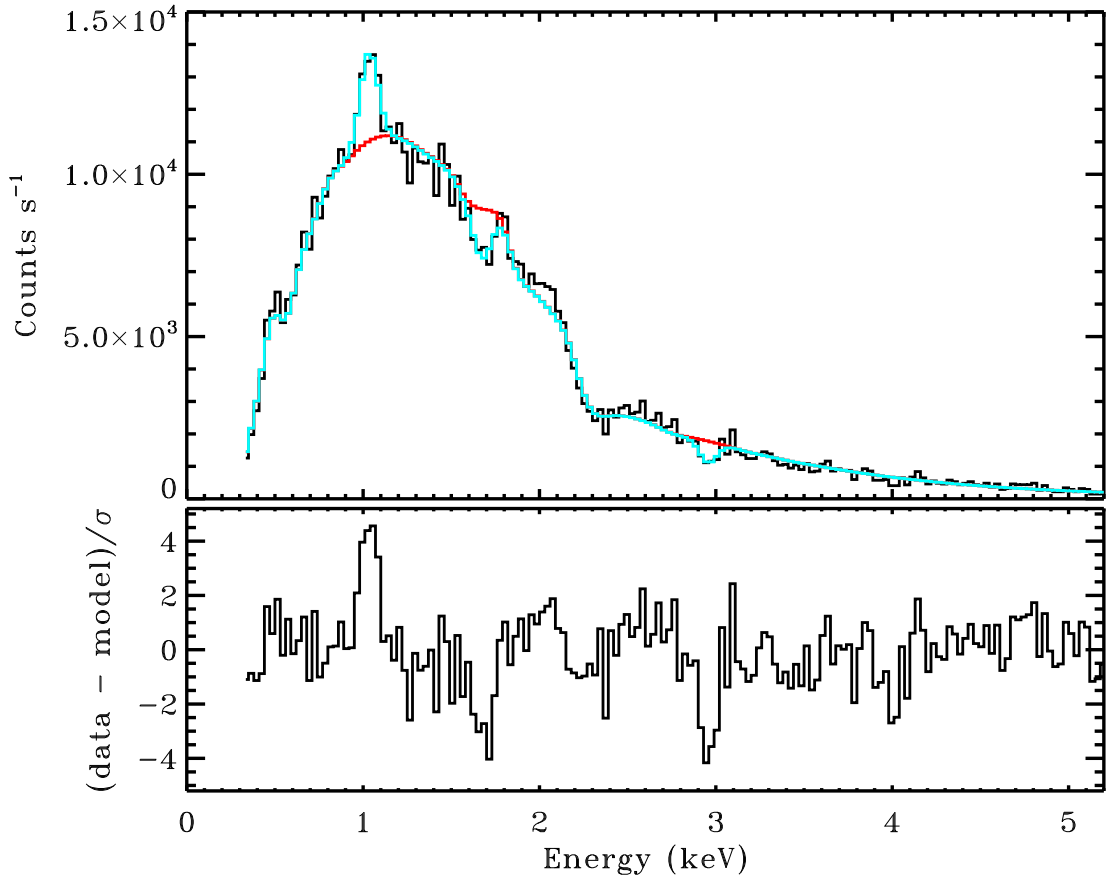}%
\end{center}
\caption{\label{fig:bp1spec} Co-added count rate spectrum of the PRE
  phases of bursts 1 and 3 (burst pair 1). The top panel shows the
  observed count rate as a function of energy. The red curve is the
  best fitting continuum model, and the cyan curve is the besting
  fitting model including the continuum and three, narrow
  gaussian line components at 1, 1.7, and 3 keV.  The fit residuals,
  in units of (data - model)/$\sigma$, for the best fitting model with
  the line normalizations set to zero, are shown in the bottom panel.
  The excess emission is clearly evident near 1 keV, and a pair of
  absorption lines are indicated near 1.7 and 3 keV.  See \S 2.3 for
  additional details.}
\end{figure*}

We fit the same model described above to both co-added spectra.
Figure \ref{fig:bp1spec} shows the spectrum from pair 1, where the red
curve (top) represents the best-fitting continuum model. It is
consistent with the individual results for bursts 1 and 3. We find for
the {\it nthcomp} component $\Gamma = 1.98 \pm 0.06$, $kT_{bb} = 0.130
\pm 0.02$ keV, and a normalization of $7.75 \pm 0.7$. For the burst
black body component we obtain $kT_{PRE} = 0.615 \pm 0.018$ keV, and
$R_{PRE} = 95.4 \pm 5$ km, which are also consistent with the
individual results for bursts 1 and 3.  We find for $N_H$ a value of
$0.17 \pm 0.03 \times 10^{22}$ cm$^{-2}$, which is consistent with the value
reported by \cite{2010ApJ...719.1807G}.

One can see in Figure \ref{fig:bp1spec} that the excess at 1 keV is
now strikingly evident, and there is also now a clearer indication for
a pair of absorption features near 1.7 and 3 keV.  We note that the
spectrum of burst 3 alone (see Figure \ref{fig:b3spec}) shows
tentative indications for these features, but they stand out more
clearly in the co-added spectrum.  To assess these features further,
we added three narrow gaussians to the model. Including the line
components dramatically improves the overall fit, and is shown as the
cyan curve in Figure \ref{fig:bp1spec}. Table \ref{table:lines}
summarizes the results of the fitted line components.  We find strong
evidence for the presence of three lines at $1.043 \pm 0.008$, $1.689
\pm 0.012$, and $2.964 \pm 0.014$ keV. The strongest, at 1 keV appears
in emission, and the others in absorption.  As an estimate of the
significance of each line we quote the ratio of the fitted line
normalization to its $1\sigma$ uncertainty.  All the lines are above
$5 \sigma$, and the 1 keV feature is at $7\sigma$.  The bottom panel
of Figure \ref{fig:bp1spec} shows the line residuals using the best
fitting model, but with the line normalizations set to zero.  The
overall spectral fit including the three line components has $\chi^2 =
279$ for 227 degrees of freedom (dof), which is formally a bit high,
however, the most substantial remaining residuals near 0.5 - 0.6 and
2.0 - 2.4 keV, are consistent with energy ranges with the most
significant unmodeled residuals in the {\it NICER} response function
\citep{2018ApJ...858L...5L}.  We emphasize that there are no
comparably strong, nor narrow, unmodeled residuals at the centroid
energies inferred for the three fitted line components.  Thus, we
think that a significant part of the excess $\chi^2$ in our fits is
likely associated with remaining uncertainties in the {\it NICER}
response function.  Fortunately, this will be testable as the {\it
  NICER} response model is refined over time.

\begin{figure*}
\begin{center}
\includegraphics[scale=0.75]{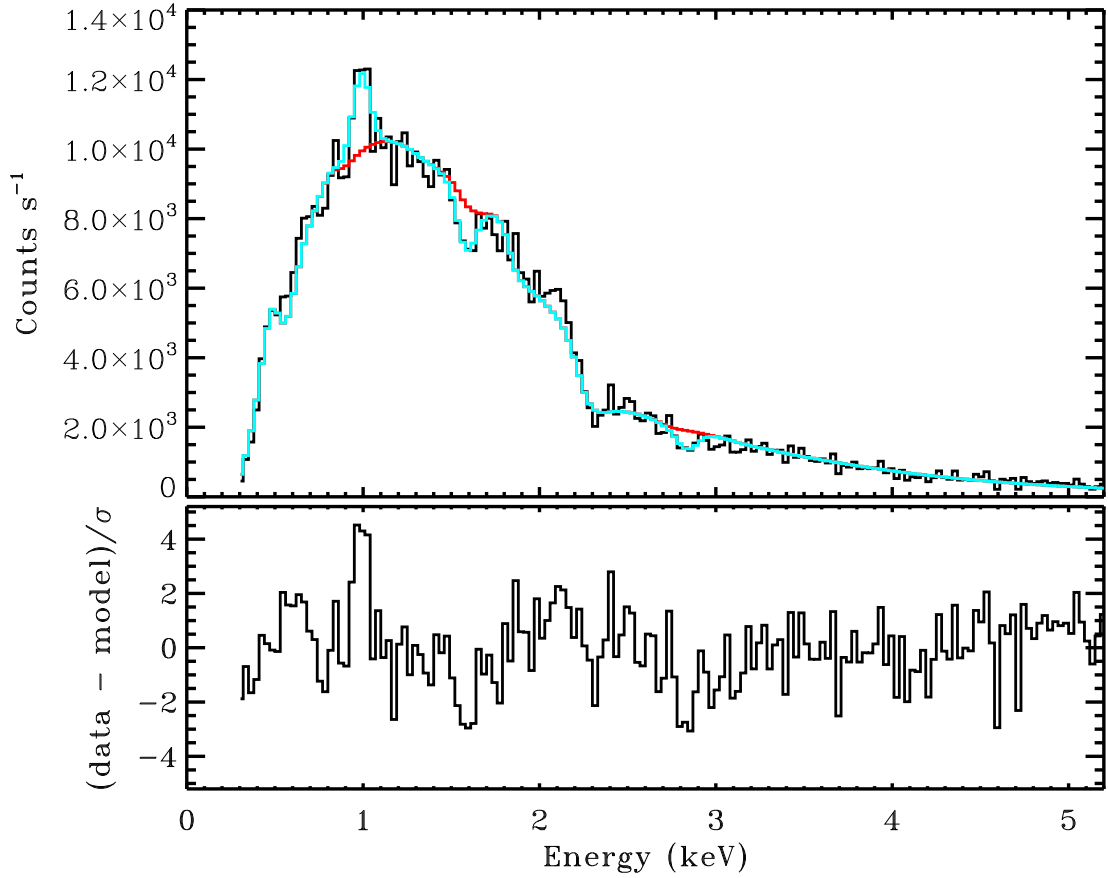}%
\end{center}
\caption{\label{fig:bp2spec} Co-added count rate spectrum of the PRE
  phases of bursts 4 and 5 (burst pair 2). Other details are the same
  as in Figure \ref{fig:bp1spec}. Features similar to those evident in
  burst pair 1 (Figure \ref{fig:bp1spec}) are also apparent in this
  spectrum. See \S 2.3 for further details.}
\end{figure*}

We carried out the same fitting procedure for burst pair 2, with the
corresponding results summarized in Figure \ref{fig:bp2spec}.
Remarkably, there is evidence for line features near the same energies
as for burst pair 1, with the emission feature near 1 keV being,
again, the clearest excess (compare Figures \ref{fig:bp1spec} and
\ref{fig:bp2spec}).  For the {\it nthcomp} continuum, in the absence
of any line components, we find $\Gamma = 1.88 \pm 0.05$, $kT_{bb} =
0.123 \pm 0.03$ keV, and a normalization of $7.82 \pm 0.6$. For the
burst black body component we obtain $kT_{PRE} = 0.675 \pm 0.026$ keV,
and $R_{PRE} = 71.6 \pm 5$ km, which are consistent with the
individual results for bursts 4 and 5.  The overall fit is of a
similar quality to that of burst pair 1, but with a slightly higher
$\chi^2 = 295$ for 227 dof. In burst pair 2 the line features appear
slightly weaker.  This is evident in the somewhat smaller $\sigma$
values (Table \ref{table:lines}). Nevertheless, they are all at values
greater than 4.7.  The fact that similar features are apparent in
both, independent sets of burst spectra is a very strong indication
that they are real, astrophysical features.

\begin{figure*}

\begin{center}
\includegraphics[scale=0.75]{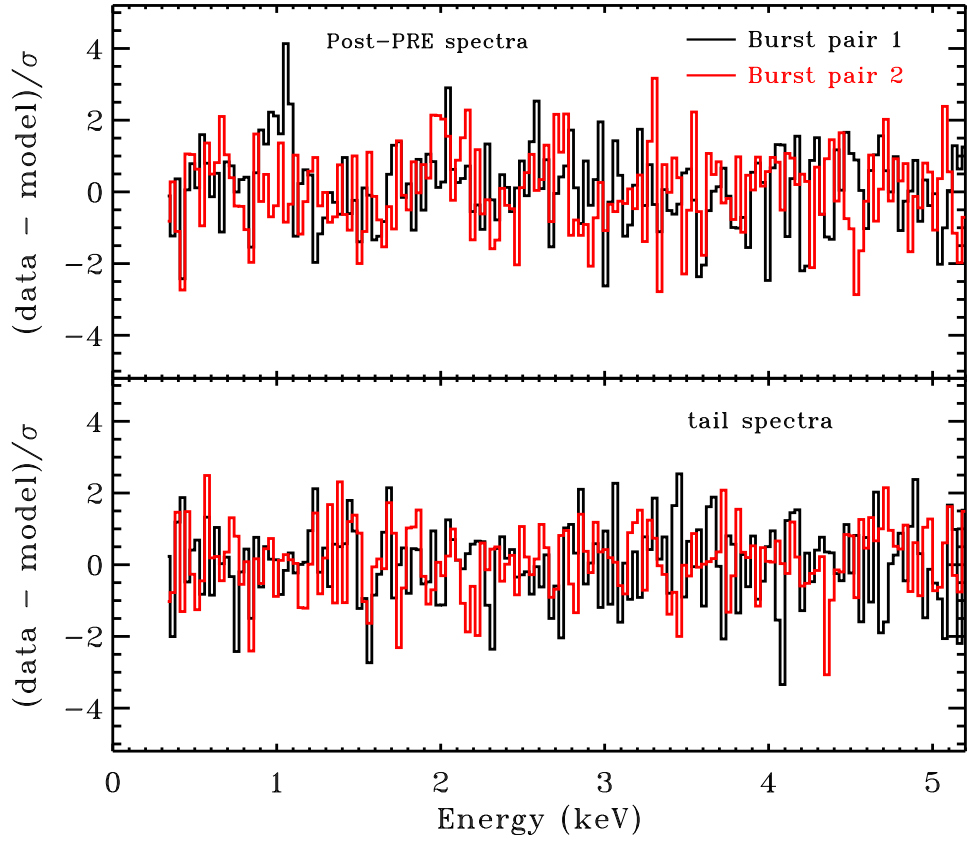}%
\end{center}
\caption{\label{fig:postpre} Fit residuals obtained from the post-PRE
  (top) and tail (bottom) intervals for both burst pairs 1 (black) and
  2 (red) are shown in units of (data - model)/$\sigma$. The
  tail spectra show no evidence for line features, and only the
  post-PRE phase for burst pair 1 shows some evidence for the 1 keV
  emission feature.}
\end{figure*}

To further explore the link between the identified line features and
the PRE phase in which they are found, we extracted spectra from
additional time intervals after the strong PRE phase to see if there
is any evidence that the lines may persist throughout the bursts. We
found no indication that the absorption lines persist beyond the
strong PRE phases described above, but we found some indication for a
persistence of the 1 keV emission line beyond the PRE phase in burst
pair 1.  As a demonstration of this we show results for spectra
accumulated for 1 s intervals {\it after} the PRE phases in which the
lines were identified, as well as 4 s intervals later in the decaying
tail. We call these the ``post-PRE'' and ``tail'' spectra,
respectively.  For consistency we co-added these spectra by burst
pairs in the same way as described above for the PRE phases, and we
fit the same model as employed for the PRE phases.  Residuals to the
fits for the post-PRE (top) and tail (bottom) spectra are shown for
both burst pairs in Figure \ref{fig:postpre}.  As is rather clear from
Figure \ref{fig:postpre} (bottom), the tail spectra show no evidence
for comparable line features like those seen in the PRE phases.  On
the other hand, for burst pair 1 there is some evidence for a
persistence of the 1 keV emission feature into the post-PRE phase (see
Figure \ref{fig:postpre}, top).  Including the 1 keV line feature to
the fit in this case gives a significance of $4.2 \sigma$, measured as
the ratio of the fitted line normalization to its $1\sigma$
uncertainty, and an equivalent width of 21 eV.  The line centroid
energy is consistent with that measured in the PRE phase.  There are
no comparable indications for persistence of the absorption features,
and the results suggest these are rather clearly linked to the strong
PRE phase.  Finally, we also analyzed a 1200 s exposure of the
persistent emission prior to burst 3.  There are no comparable line
features evident in that spectrum as well.

\begin{figure*}
\begin{center}
\includegraphics[scale=0.75]{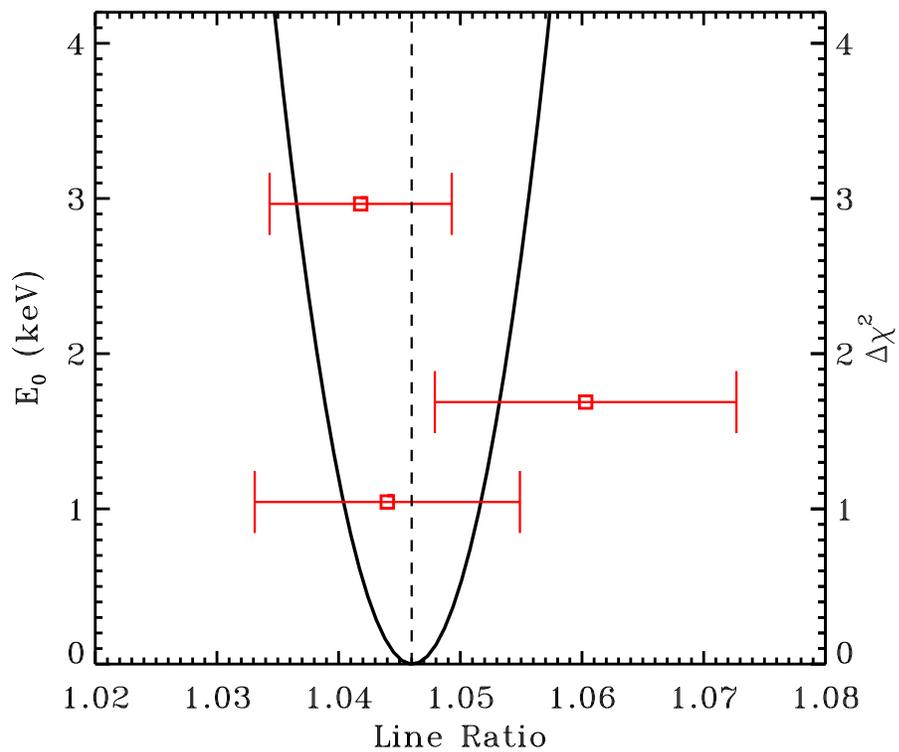}%
\end{center}
\caption{\label{fig:lineratios} Measured line energy ratios and
  uncertainties of the 1, 1.7 and 3 keV line features obtained from
  burst pairs 1 and 2 (red symbols). Also shown on the right vertical
  axis are the $\Delta\chi^2$ values for a single parameter, constant
  line ratio fit (solid black curve). The best fitting value of the
  line energy ratio, $s_0 =1.046 \pm 0.006$ ($1\sigma$, $\Delta\chi^2
  = 1$) is marked by the vertical dashed line.  A constant ratio is a
  good statistical description of the three observed line energy
  ratios. }
\end{figure*}
\begin{figure*}
\begin{center}
\includegraphics[scale=0.75]{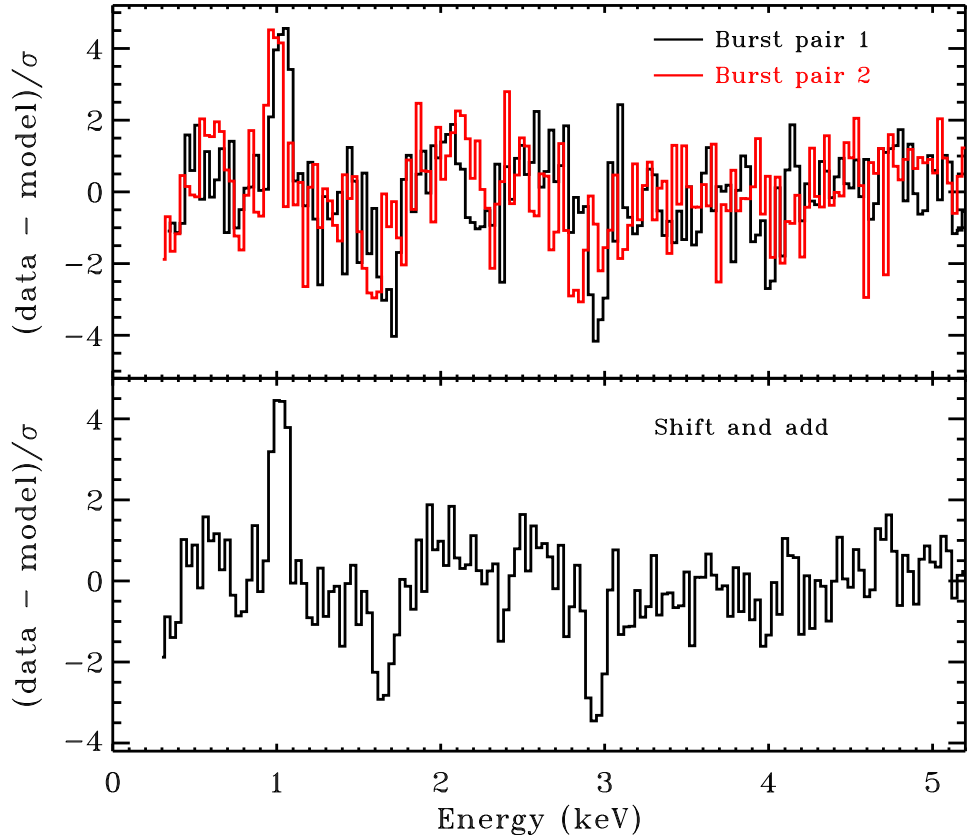}%
\end{center}
\caption{\label{fig:shiftadd} Fit residuals for both burst pairs 1
  (black) and 2 (red) are shown in units of (data - model)/$\sigma$
  (top). Here, the plotted residuals were obtained from the
  best-fitting model after setting the fitted line normalizations to
  zero.  Also shown are the average (``shift and add") residuals when
  the burst pair 2 spectrum is shifted in energy by the best fitting
  constant line ratio of 1.046 and averaged (see Figure
  \ref{fig:lineratios} and \S 2.4).}
\end{figure*}

Looking more closely at the measured line centroids and uncertainties,
there is a very significant, systematic shift when comparing the line
energies in burst pair 1 with those in pair 2 (see Table
\ref{table:lines}). All of the energy centroids measured in burst pair
2 are systematically lower.  To explore this further, we computed the
ratios of the line energy centroids in burst pair 1 to those in pair
2, and their uncertainties, and fit the values to a constant,
$s_0$. The results are shown in Figure \ref{fig:lineratios}. The line
energy ratios are statistically consistent with a constant value of
$s_0 = 1.046 \pm 0.006$, suggesting that the lines from burst pair 2
are red-shifted relative to those of pair 1.  Figure
\ref{fig:shiftadd} shows the fit residuals for both co-added spectra
using the best fitting model with the line component normalizations
set to zero.  The top panel shows together the residuals for burst
pair 1 (black) and 2 (red), and the systematic energy shift is readily
apparent.  If we shift the residuals for pair 2 by the best fit value
of $s_0 = 1.046$, and then average them with those of pair 1, we
obtain the ``shift and add'' residuals in the bottom panel.

\begin{table*}
\renewcommand{\arraystretch}{0.95}
\caption{Spectral Results for Line Components}
\label{table:lines}
\scalebox{0.95}{
\begin{tabular}{ccc}
\tableline\tableline
Parameter & Burst Pair 1 & Burst Pair 2 \\
\tableline
Centroid Energy (keV) & $1.043 \pm 0.008$  &  $1.000 \pm 0.008$  \\
Line Norm (ph)  &  $0.314 \pm 0.045$  &  $0.266 \pm 0.045$   \\
Line width (keV) & 0.06  &  0.07 \\
$\Delta\chi^2$   & 61.6  & 44.6  \\
$\sigma$  & 7.0  &  5.9  \\
equivalent width (eV) & 33  & 27 \\
\tableline
Centroid Energy (keV) & $1.689 \pm 0.012$  & $1.592 \pm 0.014$   \\
Line Norm (ph)  & $-0.125 \pm 0.023$   &  $-0.097 \pm 0.016$  \\
Line width (keV) & 0.07  &  0.07 \\
$\Delta\chi^2$   & 29.1   & 22.6  \\
$\sigma$  & 5.5   & 4.7   \\
equivalent width (eV) & $23$ & $21$ \\
\tableline
Centroid Energy (keV) & $2.964 \pm 0.014$ &  $2.846 \pm 0.0164$  \\
Line Norm (ph)  & $-0.090 \pm 0.0146$   &  $-0.079 \pm 0.0159$  \\
Line width (keV) & 0.06  & 0.07  \\
$\Delta\chi^2$   & 38.4   & 24.5  \\
$\sigma$  & 5.9   &  5.0  \\
equivalent width (eV) & $50$  & $39$ \\
\tableline
\end{tabular}}
\tablecomments{Results for gaussian line component parameters from
  fits to the co-added spectra of burst pairs 1 and 2.  In each case
  three gaussian line components have been fitted.  The widths are
  $90\%$ confidence upper limits as all the lines are unresolved. The
  value $\sigma$ is an estimate of the significance of the line, and
  is defined as the ratio of the line normalization to its uncertainty
  ($1\sigma$).}
\end{table*}

\section{Interpretation and Discussion}

We have presented strong evidence from {\it NICER} observations for
the existence of narrow spectral lines during the PRE phases of X-ray
bursts from \source{}.  The strongest feature we identify is the
emission line near 1 keV.  The properties of this feature appear
rather similar to the 1 keV emission line reported by
\cite{2013ApJ...767L..37D} in an energetic, intermediate-duration
burst from the ultracompact AMXP IGR J17062$-$6143.  \source{} is now
the second system to show such a line in emission, and it is
intriguing that both objects are ultracompact systems likely accreting
helium-rich fuel \citep{2003ApJ...595.1077C, 2018ApJ...858L..13S}.
Interestingly, \cite{2013ApJ...767L..37D} report shifts in the line
centroid energy during the IGR J17062$-$6143 burst that are at about
the same level that we find for the \source{} PRE bursts ($\approx 1.0
- 1.044$ keV).  These authors argued that the 1 keV line could be due
to Fe L shell transitions from relatively cool irradiated gas, likely
in the accretion disk. They suggested that the apparent energy shifts
could reflect contributions from different Fe charge states at
different times during the burst.  Similar processes could be at work
in \source{}.  We note that there is also the strong Ly-$\alpha$
transition of Ne X at 1.022 keV that is in the relevant energy range,
and there are several observational indications for Ne over-abundances
in some ultracompact binaries \citep{2004ApJ...608L..53S,
  2003ApJ...599..498J}.  Based on this a detection of Ne in \source{}
would not appear to be particularly surprising.  A possible scenario
then is that the emission line could be produced via reprocessing
(reflection) from the accretion disk, and the narrow line widths would
then reflect Keplerian motions in the outer disk. Using an upper limit
on the line width (HWHM) of $\sigma_E = 1.18 \times 70 = 82.6$ eV, we
would obtain $\Delta E / E \approx v/c = 82.6/1000 = 0.0826$, which
implies $R > 146 \; GM/c^2 \approx 325$ km for a 1.5 $M_{\odot}$
neutron star, which is not too dissimilar from the inferred maximum
extent of the photosphere during PRE.  This also ignores any
inclination effects.

A plausible physical interpretation for the absorption lines is that
they are formed in the super-Eddington wind generated by the PRE
bursts. \cite{2018ApJ...863...53Y} present hydrodynamical simulations
of such winds that include modern nuclear reaction networks to follow
the time-dependent burning, and composition.  They find that the
convective burning mixes the heavy element ashes of helium burning to
sufficiently low column depths such that the ashes are eventually
ejected in the wind. The ashes comprise a number of heavy elements,
including; $^{48}$Cr, $^{40}$Ca, $^{44}$Ti, $^{56}$Ni, $^{36}$Ar,
$^{32}$S, and $^{52}$Fe.  Amongst our detected absorption line
energies, we find that for the highest energy line near 3 keV there
are relatively few strong line candidates amongst those elements
listed above. A plausible identification could be the He-like lines of
S XV at 2.8839, 3.0325, and 3.1013 keV \citep{2018Galax...6...63V},
with one or more of these transitions being red- or blue-shifted to
the appropriate observed energy. We discuss likely sources of such
shifts below.  For the other absorption feature, at about 1.6 - 1.7
keV we note that there are plausible lines associated with several
charge states of Fe, including Fe XXIII, XXIV and XXV, as well as Cr
XXIV, at about the right energy. Additionally, lines of Mg XI also
fall in the appropriate range.

Alternatively, the apparent consistent energy shift between burst
pairs 1 and 2 of all three spectral features seems to argue for a
common origin, and so we briefly comment on possible origins for the 1
keV emission line in the burst-driven wind as well.  As noted by
\cite{2018ApJ...863...53Y} in their hydrodynamical simulations, the
base of the wind is initially at relatively low column depth and so
light (unburned) elements are launched first into the wind. As the
burst progresses the base of the wind moves to greater column depths,
eventually ejecting the burned ashes. Thus, composition gradients are
likely to exist in the ejected wind, and perhaps emission lines could
be formed in the outer parts of the wind. Another process which can
form emission lines is resonant scattering
\citep{2015arXiv150102776I}, and it could be that conditions are
favorable for this process to occur at some locations in the wind,
perhaps, for example, in the Ne X Ly-$\alpha$ line.  More detailed
theoretical calculations of radiative transfer in such burst-driven
winds will be needed to explore these issues further.

A difficulty with making more precise line identifications, in
addition to the need for higher spectral resolution, is the fact that
both red- and blue-shifts may be present, potentially with different
contributions in the different bursts.  Considering the gravitational
red-shift first, our spectral modeling indicates that between burst
pair 1 and 2 there was an inferred variation in the average
photospheric radius of $\approx 25$ km.  The difference in
gravitational red-shift between photons with the same rest-frame
energy emitted at two different radii $R_1$ and $R_2$, expressed as
the ratio of their observed energies, can be written as,
\begin{equation}
\frac{E_1}{E_2} = \frac{(1-2GM/c^2 R_1)^{1/2} }{(1-2GM/c^2 R_2)^{1/2}} \; ,
\end{equation}
where $E_1$ and $E_2$ are the observed energies of photons emitted
from radii $R_1$ and $R_2$, respectively.  While the continuum
spectroscopy discussed above provides good evidence that the average
radial location of the photosphere is different between burst pairs 1
and 2, it is likely that these values do not accurately reflect the
true emission radii \citep{2011A&A...527A.139S}. Nevertheless, if the
derived radii are representative, then the observed shifts are in the
correct direction, that is, the line energies for burst pair 1, with a
greater photospheric radius, are higher (less red-shifted) than those
for pair 2, with a smaller inferred radius.

\begin{figure*}
\begin{center}
\includegraphics[scale=0.75]{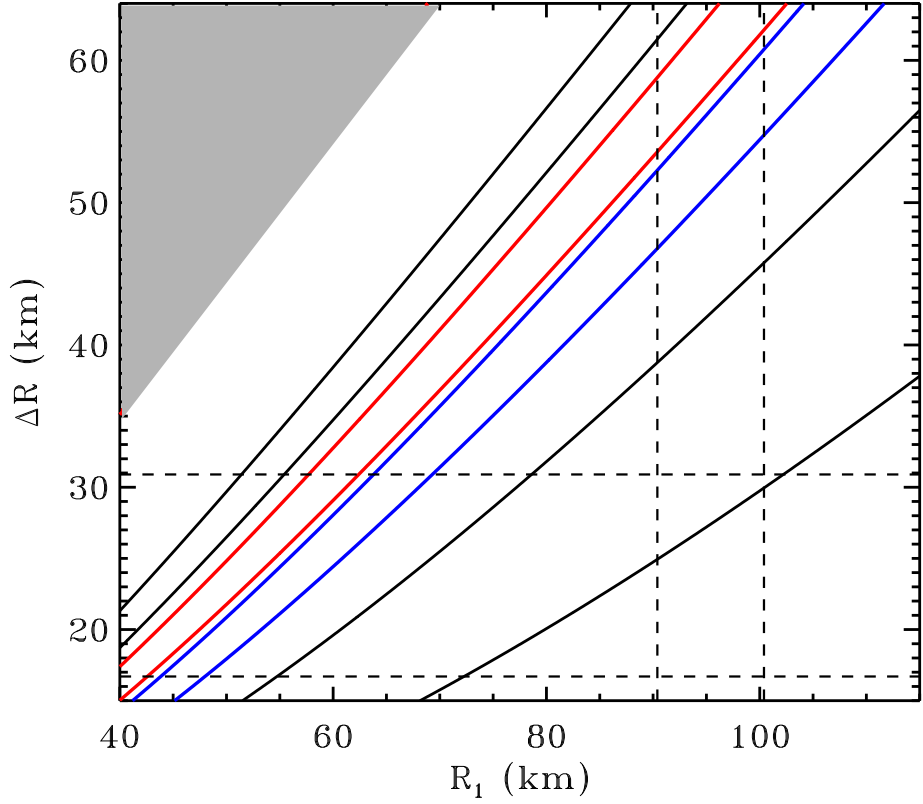}%
\end{center}
\caption{\label{fig:grshifts} Contours of constant gravitational
  red-shift, expressed as an observed energy ratio, $E_1/E_2$, are
  plotted versus $R_1$ and $\Delta R = R_1 - R_2$. Here, $R_1$ and
  $R_2$ ($R_1 > R_2$) are two radial sites of photon emission with the
  same rest-frame energy. Contours of constant line ratio consistent
  with our measured value of $1.046 \pm 0.006$ are shown for neutron
  star masses of 1.5 (red) and 2.0 (blue) M$_{\odot}$. From bottom to
  top the black curves correspond to line ratios of 1.01, 1.02, 1.06,
  and 1.08 (for a neutron star mass of 1.5 M$_{\odot}$).  Ignore the
  grey shaded region as values were not computed there. The vertical
  and horizontal dashed lines mark representative ranges of the
  photospheric radii $R_1$ and $\Delta R$ derived from co-added
  spectra of burst pairs 1 and 2 (see \S 2.3).}
\end{figure*}

Further, we can explore the assumption that {\it all} of the observed
shift is due to a gravitational red-shift and see what that implies
for the possible values of $R_1$ and $R_2$.  Figure \ref{fig:grshifts}
shows contours of constant line ratio, $E_1 / E_2$, as a function of
$R_1$ and $\Delta R = R_1 - R_2$, for 1.5 (red) and 2.0 (blue)
M$_{\odot}$ neutron stars. The red and blue contours enclose the
$1\sigma$ range of our measured line ratios for 1.5 and 2.0
M$_{\odot}$ stars, respectively.  The vertical and horizontal dashed
lines show representative ranges ($\pm 1 \sigma$) for $R_1$ and
$\Delta R$, based on the inferred photospheric radii estimated from
the co-added spectra of burst pairs 1 and 2. If the estimated radii
are at least approximately correct, then it would appear unlikely that
all of the observed shift is due to gravitational effects alone,
although we note that a heavier neutron star can more easily account
for the observed shift.

Another likely culprit to induce blue-shifts is the wind velocity
itself. Wind velocity estimates from \cite{2018ApJ...863...53Y}
suggest that the velocity is always $< 0.01 c$, which would limit a
wind-induced line ratio, $s_w < 1.01$.  Such estimates typically
ignore radiative effects such as line-driving of the wind; if these
are not significant, it would also appear that not all of the observed
shift can be attributed to the wind. However, again an apparent shift
due to this process would be in the correct direction.  The stronger
PRE bursts (pair 1) would be expected to have the stronger wind, and
thus stronger blue-shifts and higher line energies, compared to pair
2, as observed. Since both of the above processes are likely present,
and work in the same direction, it seems plausible that the observed
shifts are produced by a combination of both effects acting together.
With the present data it appears difficult to disentangle them,
however, if correlated variations in the inferred photospheric radius
and line energies with time could be accurately tracked in individual
bursts, then it might be possible to independently determine the
gravitational red-shift.

\section{Summary}

We have presented strong evidence for the presence of narrow
absorption and emission lines in {\it NICER} observations of the PRE
phases of X-ray bursts from \source{}. We find an emission line near 1
keV, and absorption features near 1.7 and 3 keV.  A significant,
systematic shift in the line energies between pairs of bursts with
different PRE strength is strongly indicative of a real physical
effect associated with the bursts, and is likely produced by a
combination of gravitational red-shift and Doppler blue-shifts
associated with the burst-driven wind.  It appears plausible that all
of the features are produced in the PRE wind, but reprocessing
(reflection) from the accretion disk could perhaps account for the 1
keV emission line.  Based on theoretical modeling of the likely
composition of the PRE-generated wind, we can tentatively identify the
3 keV absorption feature with the He-like lines of S XV.  The {\it
  NICER} data suggest a wealth of information could be gleaned from
such bursts by future observations.  For example, {\it NICER}
observations of a longer intermediate-duration burst or superburst
could perhaps track variations in the line energies as the photosphere
expands, providing a way to ``see'' the gravitational red-shift
directly. In addition, future, larger collecting area instruments,
such as {\it Strobe-X}, {\it Athena} and {\it eXTP}
\citep{2019arXiv190303035R, 2017AN....338..153B, 2019SCPMA..6229506I},
with the ability to observe with higher throughput and/or better
spectral resolution, could provide new tools to enable more precise
line identifications and perhaps use such line detections to measure
the compactness of the neutron star.

\acknowledgments

This work was supported by NASA through the {\it NICER} mission and
the Astrophysics Explorers Program. This research also made use of
data and software provided by the High Energy Astrophysics Science
Archive Research Center (HEASARC), which is a service of the
Astrophysics Science Division at NASA/GSFC and the High Energy
Astrophysics Division of the Smithsonian Astrophysical Observatory. DA
acknowledges support from the Royal Society.  SG acknowledges the
support of the Centre National d’Etudes Spatiales (CNES). GKJ
acknowledges support from the Marie Sk{\l}odowska-Curie Actions grant
no. 713683 (H2020; COFUNDPostdocDTU).

\facility{NICER, ADS, HEASARC}

\software{NICERDAS (v 004), XSPEC (Arnaud 1996)}


\bibliographystyle{aasjournal}

\bibliography{ms}

\newpage




\end{document}